\documentclass[prd, twocolumn, nofootinbib, floatfix]{revtex4-1}
\pdfoutput=1

\usepackage{hyperref}
\usepackage{amsmath}
\usepackage{graphicx}
\usepackage{dcolumn}
\usepackage{bm}
\usepackage{epsfig}
\usepackage{amssymb,latexsym,mathrsfs}
\usepackage{graphicx}
\usepackage{color}
\usepackage{multirow}
\usepackage{amssymb,latexsym,mathrsfs}
\usepackage{multirow}
\usepackage{bigints}
\usepackage{subcaption}
\usepackage{makecell}
\usepackage{textcomp}
\usepackage[title]{appendix}
\usepackage{tabularx}
\usepackage{upgreek}
\usepackage{caption}
\newcommand{\be}{\begin{equation}}
\newcommand{\ee}{\end{equation}}
\newcommand{\bea}{\begin{eqnarray}}
\newcommand{\eea}{\end{eqnarray}}
\newcommand{\iu}{{i\mkern1mu}}
\definecolor{mediumpurple}{rgb}{0.58, 0.44, 0.86}

\begin{document}

\title[Cosmology with the ET: No Slip Gravity]{Cosmology with the Einstein Telescope: No Slip Gravity Model and Redshift Specifications}

\author{Ayan Mitra$^{1}$, Jurgen Mifsud$^{2,3}$, David F. Mota$^{4}$, David Parkinson$^{2,5}$}
\affiliation{$^{1}$School of Engineering and Digital Sciences, Nazarbayev University, Nur-Sultan 010000, Kazakhstan}
\email{ayan.mitra@nu.edu.kz}
\affiliation{$^{2}$Korea Astronomy and Space Science Institute, 776 Daedeokdae--ro, Yuseong--gu, Daejeon 34055, Republic of Korea\\
$^{3}$Institute of Space Sciences and Astronomy, University of Malta, Msida, MSD 2080, Malta}
\email{jurgen.mifsud@um.edu.mt}
\affiliation{$^{4}$Institute of Theoretical Astrophysics, University of Oslo, Sem Sælands vei 13, 0371 Oslo}
\email{davidmota@astro.uio.no}
\affiliation{$^{5}$University of Science and Technology, Daejeon 34113, Republic of Korea}
\email{davidparkinson@kasi.re.kr}

\date{\today}

\begin{abstract}
The Einstein Telescope and other third generation interferometric detectors of gravitational waves are projected to be operational post $2030$. The cosmological signatures of gravitational waves would undoubtedly shed light on any departure from the current gravitational framework. We here confront a specific modified gravity model, the No Slip Gravity model, with forecast observations of gravitational waves. We compare the predicted constraints on the dark energy equation of state parameters $w_0^{}-w_a^{}$, between the modified gravity model and that of Einstein gravity. We show that the No Slip Gravity model mimics closely the constraints from the standard gravitational theory, and that the cosmological constraints are very similar.  
The use of spectroscopic redshifts, especially in the low--redshift regime, lead to significant improvements in the inferred parameter constraints. We  test how well such a prospective gravitational wave dataset would function at testing such models, and find that there are significant degeneracies between the modified gravity model parameters, and the cosmological parameters  that determine the distance, due to the gravitational wave dimming effect of the modified theory. 
\end{abstract} 
\maketitle

\section{Introduction}
\label{sec:intro}
In the current era of precision cosmology, observational probes of the expansion history and constituents of the Universe strongly rely on the so--called standard candles. Undoubtedly, Type Ia supernovae (SNe Ia) \cite{Riess:1998cb,Perlmutter:1998np,Kowalski:2008ez} have extensively been used as standard candles, since their intrinsic luminosity is assumed to be known within a certain tolerance, and therefore these could be used to determine the luminosity distance. It is well--known that gravitational waves (GWs) emerging from binary systems also encode the absolute distance information \cite{Schutz:1986gp}. The coalescence of compact binaries can be (and has been) used as standard sirens, since from the GW signal itself one would be able to measure the luminosity distance in an absolute way. These standard sirens are known to be self-calibrating, since these do not rely on a cosmic distance ladder. In order to get the redshift information of a GW event, and so place them on the luminosity distance--redshift $(D_L-z)$ relation,  an accompanying electromagnetic signal is needed (see, for instance \cite{Oguri:2016dgk,Ding:2018zrk,Abbott:2019yzh,Mukherjee:2020hyn,Yu:2020agu}, and references therein for other redshift measurement techniques in the case of dark standard sirens). Such a relation is clearly necessary  for the reconstruction of the late--time cosmological expansion of the Universe, and has also been employed to constrain various cosmological parameters of modified theories of gravity.

The era of GW astronomy began with the detection of GW150914 \cite{Abbott:2016blz} from the observation of a GW signal originating from the coalescence of a binary black hole (BBH), whereas the first detection of multi--messenger astronomy was reported with GW170817 \cite{TheLIGOScientific:2017qsa} from a GW signal emitted by a binary neutron star (BNS) inspiral accompanied by electromagnetic detections. In contrast to standard candles, the determination of a GW event's redshift is a non--trivial task, primarily because of the low resolution of the source's sky localisation which typically is of $\sim\mathcal{O}(10)\,\mathrm{deg}^2$ accuracy \cite{Fairhurst:2009tc}. On the other hand, the distance estimates from GWs are free from any external calibration requirements, which are necessary for the SNe Ia probe (the cosmic distance ladder).

The primary next generation, also known as third generation \cite{Evans:2016mbw}, GW detectors will be the ground--based Einstein Telescope (ET) \cite{Punturo:2010zz} and Cosmic Explorer (CE) \cite{3} detectors (see, for instance, \cite{Jin:2020hmc} for a comparison between the ET and CE), along with the space--based LISA/eLISA \cite{2017arXiv170200786A,AmaroSeoane:2012km} and TianQin \cite{Luo:2015ght} millihertz observatories. These GW detectors are expected to have a better sensitivity (by an order of magnitude in the strain amplitude of GWs) and a wider accessible frequency band with respect to currently available second generation detectors. The median redshift from the near future GW catalogue composed of the combined set of GW events from the planned GW detectors is envisaged to be at $z\sim2$ \cite{Congedo:2018wfn}. Clearly, such high redshift direct measurements of the luminosity distance would be clearly complementing the existing and upcoming SNe Ia measurements. In this analysis we will be focusing on the ET, although we believe that the consideration of the other third generation GW detectors would lead to interesting analyses. 

The ET is expected to make independent estimates of the several cosmological parameters, including \cite{Zhang:2018byx,Wang:2018lun,Belgacem:2019tbw}: the Hubble constant $(H_0^{})$, matter content of the Universe $(\Omega_m^0)$, spatial curvature $(\Omega_k^0)$, and the dark energy equation of state parameters ($\{w_0^{},\,w_a^{}\}$, or alternative parameterisations). Indeed, it is anticipated that more than a thousand  GW events need to be detected \cite{Cai:2016sby} (see also \cite{Liao:2017ioi} for the consideration of lensed GW events) in order to match the sensitivity of the \textit{Planck} satellite \cite{Aghanim:2018eyx}, and these envisaged to be confidently reported during the ET observation run. Although the ET would not be able to independently arrive to all the measurements of these parameters at once, the joint combination \cite{DAgostino:2019hvh,Yang:2019bpr} of the latter standard sirens data sets with the precise data sets inferred from the existing and forthcoming state--of--the--art electromagnetic probes, including measurements of the baryon acoustic oscillations (BAO) and of the cosmic microwave background (CMB), would significantly enhance our knowledge on the dynamics of the Universe. Furthermore, the opportunity of observing black holes back to a much earlier epoch of the Universe could allow us to observe the remnants of the first stars, and to explore the dark ages,  during which proto-galaxies and large--scale structure emerged.


A number of recent works (see, for instance, \cite{6,Yang:2019vni,Zhang:2019ple,Wolf:2019hun,Yang:2020wby,Bachega:2019fki,Li:2019ajo}, and references therein) have illustrated the strength of GW detections by their ability to constrain different dark energy models. Next generation GW detections also gives us the scope to perform tests on theories of modified gravity by confronting the modified propagation of GWs across cosmological distances. This is possible because of their higher source redshift and lesser calibration requirements. Modified theories of gravity are normally characterised by different evolution of scalar as well as tensor perturbations \cite{Nishizawa:2017nef}. Consequently, any deviation from Einstein gravity could be parametrised in the propagation equation of GW by introducing new parameters related to (for example) the propagation speed of GWs, a friction term which dilutes the amplitude of GWs, graviton mass, or an energy source term. GW probes, particularly the upcoming detectors, have been shown (see, for instance, \cite{Sathyaprakash:2009xt,Monitor:2017mdv,Wang:2018lun,Yang:2019vni,Yang:2018bdf,Zhang:2018byx,Du:2018tia,Lagos:2019kds,Belgacem:2019zzu, Zhang:2019ple,Bachega:2019fki,Yang:2019bpr,Chen:2020lzc,Sharma:2020btq,Mastrogiovanni:2020gua} and references therein) to be able to help shed light on deviations from Einstein gravity. 

In this paper, we will present a comparative study of the cosmological dark energy parameter constraints we will be expecting from the upcoming GW observations between the existing standard dark energy models and modified gravity models. We also investigate how tests of the models may be confused by degeneracies between the modified theory predictions, and parameters that control the distance. In section \ref{sec:et} we discuss the proposed third generation GW detectors, while in section \ref{sec:theory} we briefly review the theoretical framework of modified GW propagation. In section \ref{sec:data} we illustrate the data and methodology which will be implemented in section \ref{sec:res}. We draw our final conclusions and prospective lines of research in section \ref{sec:con}.

\begin{figure}
    \includegraphics[width=\columnwidth]{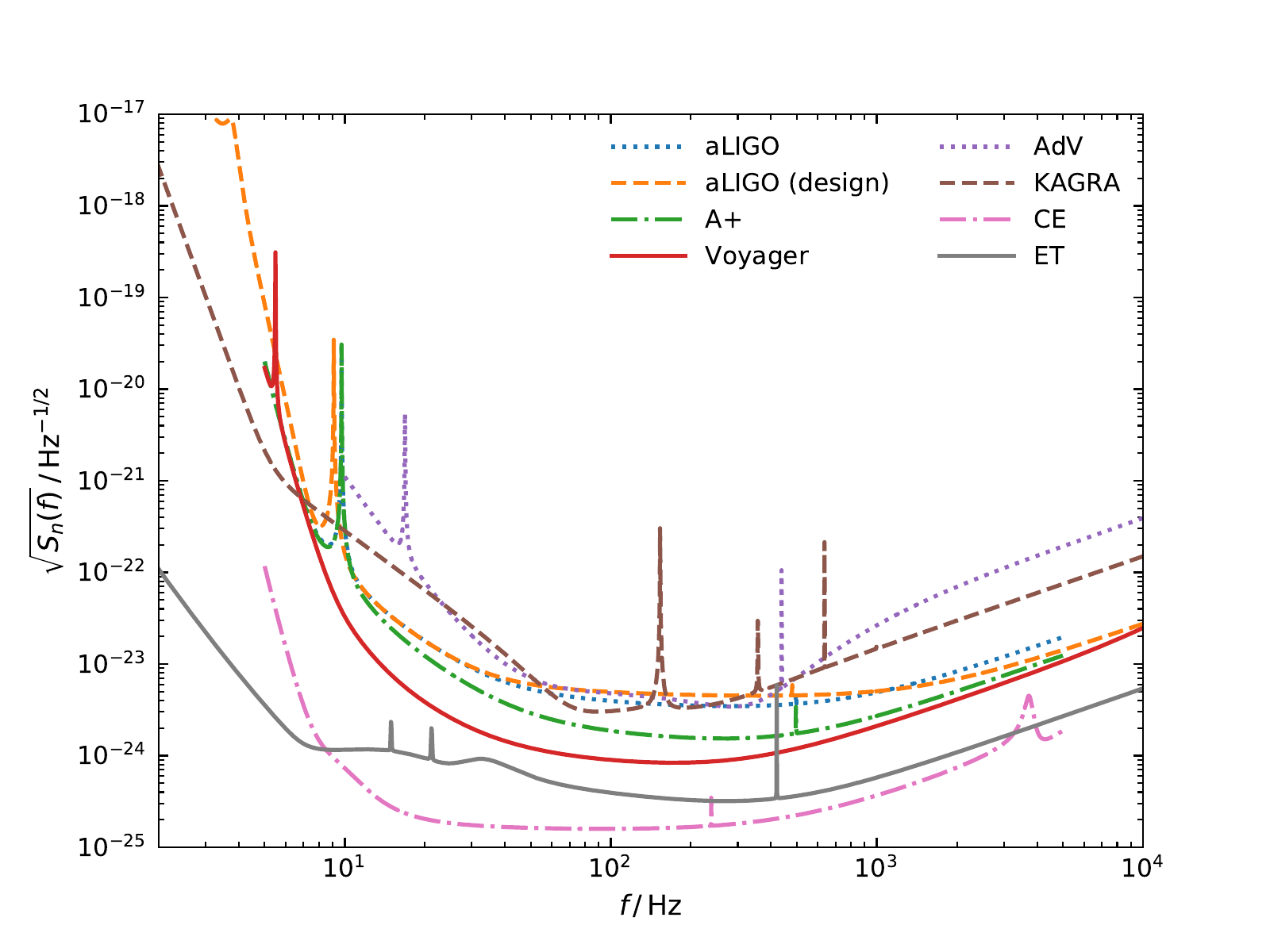}
    \caption{Sensitivity evolution of current and proposed GW interferometric detectors.} 
    \label{fig:noises}
\end{figure} 

\section{Third Generation GW Detectors}
\label{sec:et}

The achieved sensitivity by the first generation of interferometric detectors (LIGO \cite{Abbott:2007kv}, Virgo \cite{Acernese:2008zza},
GEO 600 \cite{Grote:2008zz} and TAMA \cite{Takahashi:2004rwa}) was mainly limited by shot noise, mirror thermal noise and seismic noise, while for the second generation GW detectors, such as Advanced LIGO (aLIGO) \cite{Harry:2010zz}, Advanced Virgo (AdV) \cite{TheVirgo:2014hva}, KAGRA \cite{Somiya:2011np,Aso:2013eba}, and LIGO-India \cite{Unnikrishnan:2013qwa} additional fundamental noise sources will play a role towards the low--frequency end of the detection band. As expected, the latter noise sources will be more prominent in third generation GW detectors \cite{Hild:2010id,Punturo:2010zzb,Huttner:2016dzl}, particularly due to the fact that the main aim of these detectors is to probe the low--frequency band; as low as a few Hz \cite{Hild:2009ns}. This low--frequency range is one of the main driving forces of third generation GW detectors, since it encapsulates some rich information on the cosmological evolution of the Universe (see, for instance, \cite{Punturo:2010zza,Sathyaprakash:2012jk,Srivastava:2019fcb,Maggiore:2019uih,Zhang:2019ple,Yang:2019vni,Yang:2019bpr,Bachega:2019fki,Sathyaprakash:2009xt,Chen:2020lzc,Sharma:2020btq}, and references therein).

\begin{table}
    \setlength\extrarowheight{2.09pt}
    \begin{center}
        \begin{tabular}{ c c c } 
            \hline
            $a_t$ & $\tau$ & $\mu$ \\[.3em]
            \hline
            \multirow{3}{*}{0.5} & 1.5 & 0.1 \\ 
            & 1.0 & 0.1 \\ 
            & 1.0 & 0.2 \\ 
            \hline
        \end{tabular}
    \end{center}
    \caption{The three different  combinations of adopted parameter values for $a_t,\,\tau$ and $\mu$ \cite{5}.}
    \label{t0}
\end{table}

We should also remark that this young field of observational astrophysics is constantly being enhanced by technological improvements. Indeed, the current GW detectors are expected to be upgraded to Advanced LIGO plus (A+) \cite{LIGOScientific:2019vkc} (possibly to LIGO Voyager \cite{Adhikari:2019zpy}), Advanced Virgo Plus (AdV+) \cite{Michimura:2019cvl}, and KAGRA+ \cite{Akutsu:2020zlw}. Figure~\ref{fig:noises} illustrates the  amplitude spectral densities \cite{ETnoises} of the ET along with some of the mentioned second and third generation GW detectors as a function of frequency. As clearly indicated in this figure, the ET would be able to probe a considerably wide range of frequencies with significantly good sensitivity with respect to the upcoming GW detectors.

As already mentioned, we will be considering the ET specifications \cite{Punturo:2010zz} for our analyses, which is a proposed third generation ground--based interferometric detector expected to be fully operational in early $2030$s. It will be observing GWs emanated from BBH mergers up to $z\lesssim20$, the coalescence of BNS systems up to $z\lesssim2$, as well as from neutron star--black hole (NSBH) inspirals up to $z\lesssim8$. The ET is envisaged to detect $\mathcal{O}(10^3-10^7)$ BNS events per year with signal--to--noise--ratios (SNRs) ranging up to $\sim100$ \cite{Messenger:2011gi,Abadie:2010cf,Sathyaprakash:2009xt}, with a fraction of these events having an electromagnetic afterglow \cite{6,Sathyaprakash:2009xt}. The current second generation kilometer--scale GW detectors target frequency windows in the range of $\sim20-2000\,\mathrm{Hz}$, while next generation interferometers will be able to probe frequencies as low as $\sim 1$ Hz \cite{Tamanini:2016zlh}. The frequency range is important since it determines the masses of compact objects that the GW detector could observe.

GW interferometric detectors are sensitive to the relative difference between two distances, the so--called strain $h(t)$, with $t$ being the cosmic time. It is well--known that GWs are characterised by a second rank tensor $h_{\alpha\beta}$, having only two independent components $h_+$ and $h_\times$ in the transverse--traceless gauge, since the non---zero components are $h_{xx}=-h_{yy}=h_+$ and $h_{xy}=h_{yx}=h_\times$. The response function of a given GW detector is given by
\begin{equation}
    h(t)=F_+(\vartheta,\,\varphi,\,\psi)h_+(t)+F_\times(\vartheta,\,\varphi,\,\psi)h_\times(t)\;,
\end{equation}
where $F_+$ and $F_\times$ are the detector's antenna pattern functions, $(\vartheta,\varphi)$ are the angles describing the location of the source on the sky, and $\psi$ is the polarisation angle. The antenna pattern functions of the ET are given by \cite{6}
\begin{align}
    F_+^{(1)}(\vartheta,\,\varphi,\,\psi)=&
    \begin{aligned}[t]
        &\frac{\sqrt{3}}{2}\Bigg[\frac{1}{2}\left(1+\cos^2(\vartheta)\right)\cos\left(2\varphi\right)\cos\left(2\psi\right)\\
        &\;\;\;\;\;\;\;-\cos(\vartheta)\sin\left(2\varphi\right)\sin\left(2\psi\right)\Bigg]\;,
    \end{aligned}\\
    F_\times^{(1)}(\vartheta,\,\varphi,\,\psi)=&
    \begin{aligned}[t]
        &\frac{\sqrt{3}}{2}\Bigg[\frac{1}{2}\left(1+\cos^2(\vartheta)\right)\cos\left(2\varphi\right)\sin\left(2\psi\right)\\
        &\;\;\;\;\;\;\;+\cos(\vartheta)\sin\left(2\varphi\right)\cos\left(2\psi\right)\Bigg]\;,
    \end{aligned}
\end{align}
with the remaining two antenna pattern functions being $F_{+,\times}^{(2)}(\vartheta,\,\varphi,\,\psi)=F_{+,\times}^{(1)}(\vartheta,\,\varphi+2\pi/3,\,\psi)$ and $F_{+,\times}^{(3)}(\vartheta,\,\varphi,\,\psi)=F_{+,\times}^{(1)}(\vartheta,\,\varphi+4\pi/3,\,\psi)$. We remark that the latter two antenna pattern functions follow from the equilateral triangle design of the interferometric detector.

Following the stationary--phase approximation \cite{6} which applies due to the adiabatic evolution of the inspiral's wave frequency, we arrive at the Fourier transform $\widehat{\mathcal{H}}(f)$ of the time--domain waveform $h(t)$,
\begin{equation}
    \widehat{\mathcal{H}}(f)=\mathcal{A}f^{-7/6}\exp{\left[\iu\left(2\pi f t_0^{}-\pi/4+2\varPsi(f/2)-\varPhi_{(2,0)^{}}\right)\right]}\;,
\end{equation}
where $\mathcal{A}$ is the Fourier transform amplitude, given by
\begin{equation}
\begin{split}
    \mathcal{A}=&\frac{1}{D_{L,\mathrm{GW}}}\sqrt{F_+^2\left(1+\cos^2(\omega)\right)^2+4F_\times^2\cos^2(\omega)}\\
    &\hspace{.5cm}\times\sqrt{5\pi/96}\pi^{-7/6}\mathcal{M}_c^{5/6}\;.
\end{split}
\end{equation}
In the above, we are considering a coalescing binary system located at a characteristic luminosity distance $D_{L,\mathrm{GW}}$, having a total mass of $M=m_1+m_2$, with component masses $m_1$ and $m_2$. The associated observed chirp mass is denoted by $\mathcal{M}_c=(1+z)M\chi^{3/5}$, with $\chi=m_1m_2/M^2$ being the symmetric mass ratio. Moreover, the constant $t_0^{}$ denotes the epoch of the merger, while $\omega$ is the angle of inclination of the binary's orbital angular momentum with the line--of--sight. The introduced functions are specified by
\begin{eqnarray}
    \varPsi(f)&=&-\varPsi_0^{}+\frac{3}{256\chi}\sum_{i=0}^7\varPsi_i^{}(2\pi Mf)^{i/3}\;,\\[1em]
    \varPhi_{(2,0)}&=&\arctan\left(-\frac{2\cos(\omega)F_\times}{(1+\cos^2(\omega))F_+}\right)\;,
\end{eqnarray}
where the parameters $\varPsi_i^{}$ are reported in \cite{Sathyaprakash:2009xs}. 

\section{Theory}
\label{sec:theory}

In Einstein gravity, the linearised evolution equation of GWs propagating in a spatially--flat  Friedmann--Lema\^{i}tre--Robertson--Walker (FLRW) background is given by
\begin{equation}
     h''_A+ 2{ \cal H}\, h'_A+k^2h_A\,=\,\Pi_A\;,
\end{equation}
where the primes indicate the derivatives with respect to conformal time $\eta$, $A=[\times,\,+]$ corresponds to the two polarisation states, $h$ are the Fourier modes of the GW's strain amplitude, $\mathcal{H} = a'/a$ is the conformal Hubble parameter such that $a=(1+z)^{-1}$ is the scale factor, and the term on the right hand side is the source term related to the anisotropic stress tensor. However, in the case of a slightly more generic theory of modified gravity, the propagation equation of GWs changes to 
\begin{equation}
     h''_A+ 2{ \cal H}\left[1-\delta(\eta)\right]\, h'_A+k^2h_A\,=0\;,
\end{equation}
where we have retained any deviation from the standard prediction by the function $\delta(\eta)$. It modifies the friction term in the propagation equation of GWs over a cosmological background, and thus describes the effect of propagation of the modified GWs (we will present the parametrisation of $\delta(\eta)$ later). The modified middle term is important as it affects the amplitude of GWs propagating across cosmological distances, and hence the definition of the GW luminosity distance. In Einstein gravity we have that $\delta(\eta)=0$, whereas in a number of modified gravitational theories $\delta(\eta)$ is directly linked with the effective Planck mass.

\begin{figure}
    \includegraphics[width=\columnwidth]{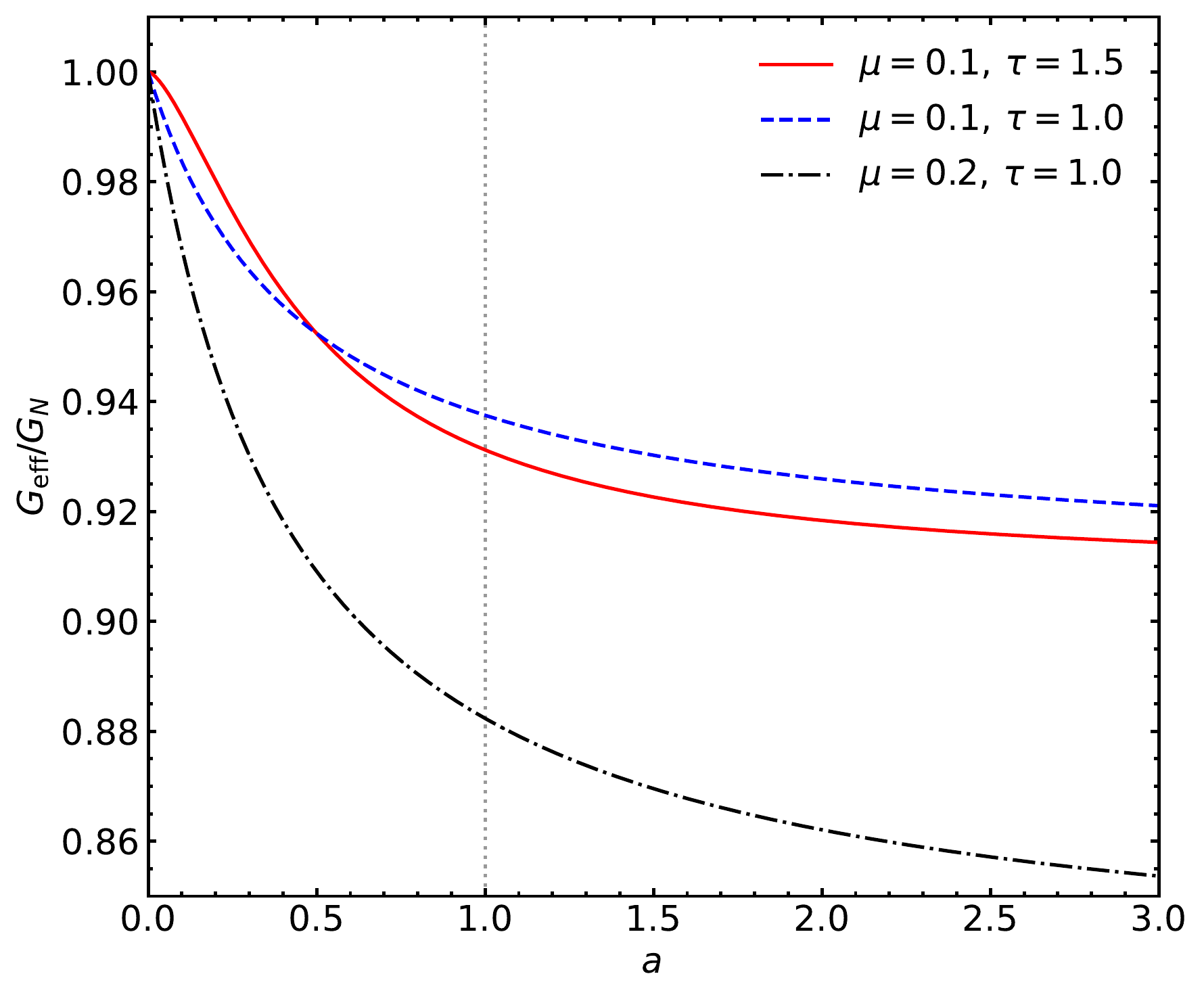}
    \caption{The comparison of Planck masses via $G_\mathrm{eff}/G_N=G_\mathrm{matter}=G_\mathrm{light}$, for the indicated values of $\mu$ and $\tau$, with fixed $a_t=0.5$. We can verify that at very early--times $(a\rightarrow0)$, $M_*\simeq m_p$.} 
    \label{planck}
\end{figure} 

In the following analysis, we will be considering the inferred dark energy parameter constraints from upcoming detections of standard sirens, using a specific modified gravity model; the No Slip Gravity model \cite{5}. No Slip Gravity is a special subclass model of the Horndeski gravitational framework \cite{Horndeski:1974wa,Deffayet:2011gz}, which is well--known to be the most general scalar--tensor theory having second--order field equations in four dimensions. This framework encompasses \cite{15}, for instance, $f(R)$ models \cite{DeFelice:2010aj}, quintessence \cite{Tsujikawa:2013fta},  the Brans--Dicke model \cite{Brans:1961sx}, and covariant Galileons \cite{Deffayet:2009wt}.

The  No Slip Gravity model is advantageous to study in the sense that gravitational waves propagate at the same  speed of light. Recent results from the binary neutron star merger GW170817/GRB 170817A  have shown that the  speed of propagation of GWs ($c_\mathrm{GW}$) is in an excellent agreement with the speed of light ($c$), such that $|(c_\mathrm{GW}-c)/c|\lesssim\mathcal{O}(10^{-15})$ \cite{12}.  The No Slip Gravity model is therefore a viable model which naturally satisfies this requirement, as opposed to a number of well--known modified theories of gravity which were adversely affected by this measurement (see, for instance, \cite{Ezquiaga:2017ekz,Sakstein:2017xjx,Amendola:2017orw,Crisostomi:2017pjs,Baker:2017hug,Ezquiaga:2018btd,Kase:2018aps}). 

\begin{figure}
    \includegraphics[width=\columnwidth]{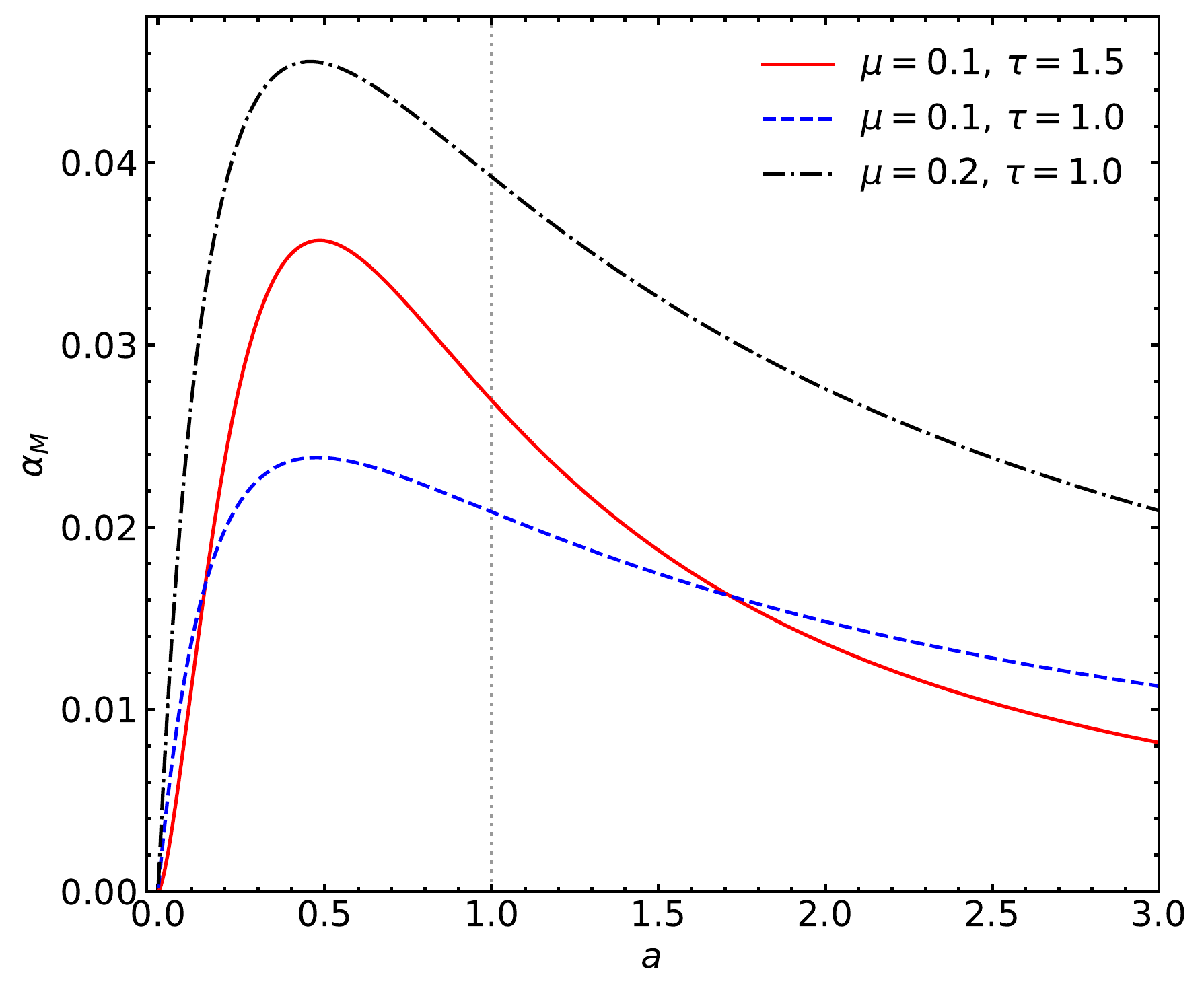}
    \caption{Comparison plot of $\alpha_M$, for the indicated values of $\mu$ and $\tau$, with fixed $a_t=0.5$. The grey vertical line indicates the corresponding $\alpha_{M}^0$ values.} 
    \label{alpha}
\end{figure}

The gravitational effect on matter and photons can be analysed via the modified Poisson equations which relate the time--time metric potential with the space--space metric potential. The growth of cosmic structure is governed by the gravitational strength $G_\mathrm{matter}$, while the deflection of light is characterised by the gravitational strength $G_\mathrm{light}$. The offset between the latter gravitational strengths is referred to as the gravitational slip parameter, defined by
\begin{equation}
    \Bar{\eta}=\frac{G_\mathrm{matter}}{G_\mathrm{light}}\;,
\end{equation}
such that $\Bar{\eta}=1$ corresponds to vanishing slip, which holds in the case of the concordance model of cosmology. In this modified gravity model, we have a simple relationship between the Planck mass running parameter $\alpha_M$, and the kinetic braiding parameter $\alpha_B$ (see \cite{Bellini:2014fua} for further information on the Horndeski property functions $\alpha_{M,B}$). Indeed, the no slip condition is specified by $\alpha_B=-2\alpha_M$, which then determines the ratio between the constant Planck mass $(m_p)$ in Einstein gravity, and the effective time--dependent Planck mass in modified gravity $M_\ast$, which is given by
\begin{equation}
    G_\mathrm{matter}=G_\mathrm{light}=\frac{m_p^2}{M_*^2}\;.
\end{equation}
We should remark that a consequence of the stability conditions within this framework, the gravitational strength is found to be diluted with respect to the standard prediction, leading to weaker gravity. This atypical feature of scalar--tensor theories of gravity arises from the fact that the non--null kinetic braiding parameter mixes the scalar sector into the tensor sector, and such a feature could address possible anomalies in growth of structure observations \cite{Hildebrandt:2016iqg,Joudaki:2017zdt,Abbott:2017wau,Troxel:2017xyo}.

When considering the propagation of GWs, it is essential that we infer the luminosity distance of the source, $D_{L,\mathrm{GW}}$. The standard luminosity distance for electromagnetic sources will be denoted by $D_{L,\mathrm{GR}}$, such that 
\begin{equation}
D_{L,\mathrm{GR}}(z)=c(1+z)\int^{z}_{0}\frac{\mathrm{d}z'}{H(z')}\;, \label{h0}
\end{equation}
with
\begin{equation}
H^2(z)=H_0^2\left[\frac{\Omega_m^0}{a^{3}}+\frac{(1-\Omega_m^0)}{a^3}\left(a^{-3(w_0+w_a)}e^{3[w_a(a-1)]}
\right)\right]\;, \label{h1}
\end{equation}
where we recall that $H_0$ and $\Omega_m^0$ denote the Hubble constant and the current matter density fraction, respectively. For the dark energy parameter choice of ${(w_0,\,w_a)}={(-1,\,0)}$, i.e. the concordance model of cosmology, the above relation for $H(z)$ reads as follows 
\begin{equation}
H^2(z)=H_0^2\left[\Omega_m^0(1+z)^{3}+(1-\Omega_m^0)\right]\;, 
\end{equation}
for a spatially--flat FLRW metric. In the rest of the paper, unless explicitly mentioned, $D_L$ will be denoting $D_{L,\mathrm{GR}}$.

    \begin{figure}
        \includegraphics[width=\columnwidth]{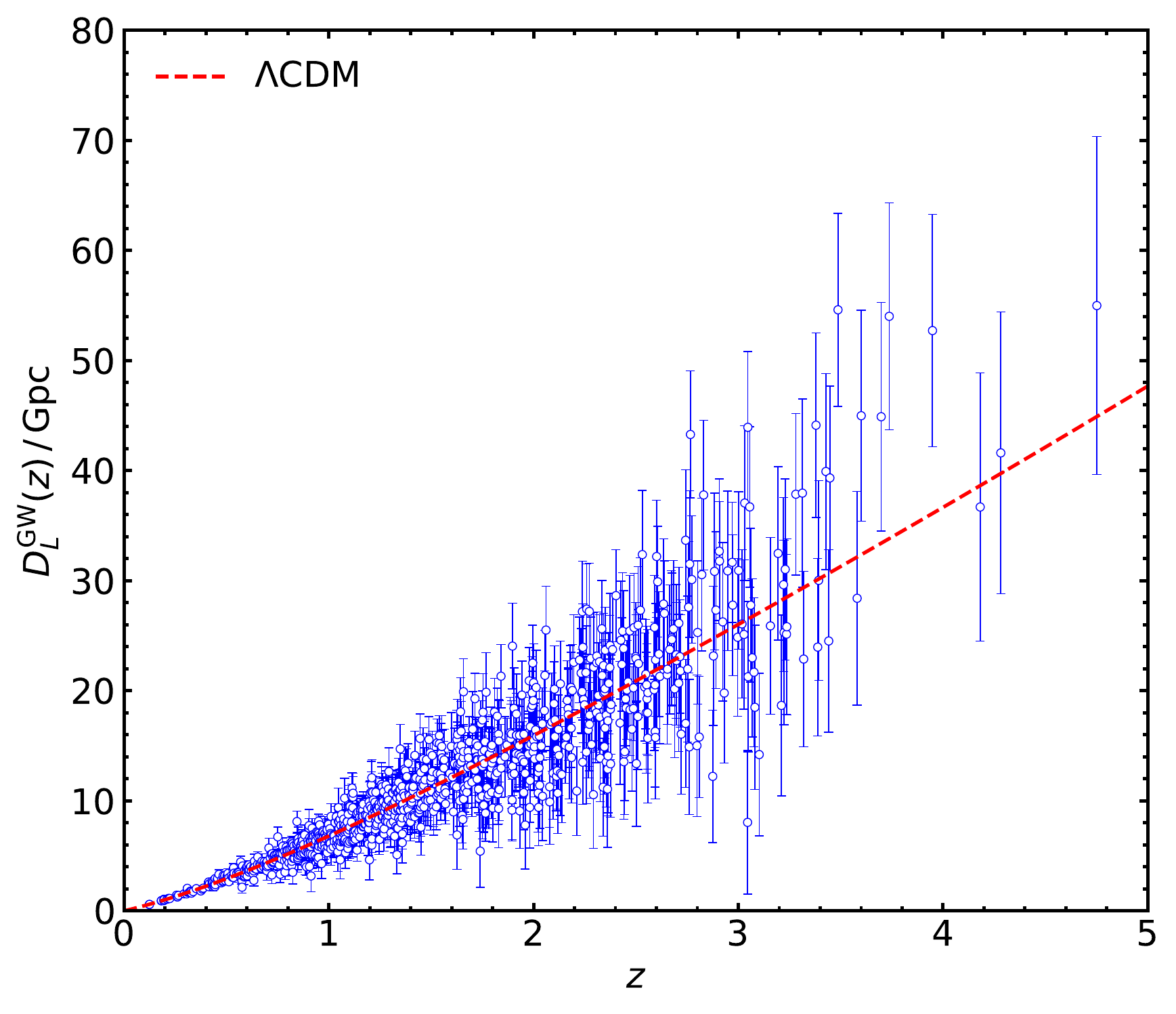}
        \caption{Luminosity distance distribution of the GW data as a function of redshift. There are a total of $1000$ GW candidates.} 
        \label{fig:DL_GW}
    \end{figure} 

The relationship between the GW standard siren luminosity distance and the photon standard candle luminosity distance, is given by \cite{eq15a,eq15b,Nishizawa:2017nef,13,14},
\begin{equation}
     \frac{D_{L,\mathrm{GW}}(z)}{D_{L,\mathrm{GR}}(z)}=\exp{\left\{-\int_0^z\frac{\mathrm{d}z'}{1+z'}\delta(z')\right\}}\;.
     \label{e4}
\end{equation}
We will now consider the determination of $\{M_*,\,\alpha_M\}$ by adopting the parametrisations as reported in \cite{5}, which is explicitly given by
\begin{equation}\label{9}
    \left(\frac{m_p}{M_*}\right)^{-2}=1+\frac{\mu}{1+\left( a/a_t\right)^{-\tau}}\;,
\end{equation}
where $\mu$ is the amplitude of the transition from the early Universe to the asymptotic future, and $a_t$ is the scale factor when this occurs, with $0<\tau\leq3/2$ being its rapidity. In Figure~\ref{planck} we illustrate three comparisons of the Planck mass with the time--dependent effective Planck mass according to the parameter values as specified in Table~\ref{t0}. We can verify that, asymptotically the effective running Planck mass tends to the constant Planck mass; this is justified since in the early Universe they are expected to be identical.

By using the fact that $\alpha_M=\mathrm{d}\ln M_\ast^2/\mathrm{d}\ln a$, we get to the following parametrisation of the Planck mass running parameter
\begin{equation}
   \alpha_M=\left[1+\frac{\mu}{1+e^{-\tau(\ln a-\ln a_t)}}\right]^{-1} 
\frac{\tau\mu e^{-\tau(\ln a-\ln a_t)}}{\left[1+e^{-\tau(\ln a-\ln a_t)}\right]^2}\;.  \label{10}  
\end{equation}
The redshift evolution of $\alpha_M$ is depicted in Figure~\ref{alpha}, for the specified parameter values listed in Table~\ref{t0}. In order to solve equation (\ref{e4}), we need to find an expression for $\delta(z)$. \cite{16} has provided a generic parametric form of $\delta(z)$, suitable for most of the modified gravity models,
\begin{equation}
\label{delta2}
    \delta(z) = \frac{n(1-\zeta_0)}{1-\zeta_0+\zeta_0(1+z)^n}\;,
\end{equation} 
where $\zeta_0$ and $n$ for Horndeski specific models are defined by
\begin{eqnarray} 
 \zeta_0&=&\lim_{z \to \infty}  \frac{M^*(0)}{M^*(z)}\;,\label{z0}\\[1em]
 n&\simeq&\frac{\alpha_{M0}}{2(\zeta_0 - 1)}\;.\label{z1}
\end{eqnarray}
We can see that at early--times ($z\rightarrow\infty$) we recover Einstein gravity with $\delta(z)\rightarrow0$, while at late--times ($z\ll1$), $\delta(z)\simeq n(1-\zeta_0)$. $\zeta_0=1$ corresponds to the standard prediction, thus the two luminosity distances are identical, $D_{L,\mathrm{GW}} =D_{L,\mathrm{GR}}$.

\section{Analysis}
\label{sec:data}

\subsection{Data}
We will present a comparison of our results obtained from using the luminosity distance estimates using the modified gravity model, to the ones derived from assuming a standard electromagnetic source luminosity distance. For computing the latter case, we used the simulation data of \cite{17}. We refer the reader to this work for further details on the characteristics of the data. 

\subsection{Errors}
\label{sec:errors}

For Fisher matrix analysis \cite{nla.cat-vn81100,Cutler:1994ys,Berti:2004bd}, we need to model the systematic errors and account for the cosmological uncertainties propagating through the luminosity distance measurement. The uncertainties ($\sigma$) in luminosity distance measurement error $\langle\delta D_L \rangle$ is composed of
\begin{equation}
    \langle\sigma  \rangle^2=\langle \sigma_\mathrm{photo-z} \rangle^2+\langle \sigma_{WL} \rangle^2+\langle \sigma_{I} \rangle^2+\langle \sigma_{P} \rangle^2\;,  
\end{equation}
where the four terms on the right--hand side stand for the photometric  redshift measurement error, weak lensing error, instrumental error and the peculiar velocity error.

\begin{itemize}
    \item \textbf{Redshift error :} Since most of distant binaries will have photometric redshift, it is essential to account for the photometric redshift measurement error. It is modelled by \footnote{The spectroscopic redshift measurement error is neglected here; \cite{Congedo:2018wfn} state that the nominal requirements from EUCLID/DESI is of $\sigma_\mathrm{spec}=0.001$.}\cite{Tamanini:2016zlh,19,20}
    \begin{equation}
        \sigma_\mathrm{photo-z} = \left( \frac{\partial D_L}{\partial z}\right)\left[0.03(1+z)\right]\;. 
    \end{equation}
    It is vital, that of all the sources which will be detected, the fraction of counterparts identifiable with the availability of spectroscopic redshift should have a significant effect on the parameter estimation and the improvement of the constraints. For the spectroscopic redshift sources, we assume a flat error of $0.001$ \cite{Congedo:2018wfn}. Even with current ongoing large scale SNe surveys, mitigating the systematic errors originating from high--redshift photometric samples is a challenge in itself. Assuming there will be vast improvement in observing capabilities in the next one--to--two decades in redshift measurements, we will present a comparison of the results based on scenarios, of different level of expected detections of the spectroscopic redshift.\\    
    \begin{table}
        \setlength\extrarowheight{2.09pt}
        \begin{center}
            \begin{tabular}{ c c c c }
                \hline
                $\alpha_{M}^0$ & $\zeta_0$ & n \\[.3em]
                \hline
                0.0270 & 1.0363 & 0.3715 \\ 
                0.0208 & 1.0328 & 0.3176 \\ 
                0.0392 & 1.0646 & 0.3036 \\ 
                \hline
            \end{tabular}
        \end{center}
        \caption{We here summarise the fit parameters which are necessary for the computation of the luminosity distances in our modified gravity model. These have been inferred from equation (\ref{e4}) along with the specified values in Table~\ref{t0}. }
        \label{T1}
    \end{table}
    \item \textbf{Instrumental error :} 
    The combined SNR for the proposed ET's network of three independent interferometers is given by
    \begin{equation}
        \rho=\sqrt{\sum_{i=1}^3\left(\rho^{(i)}\right)^2}\;,
    \end{equation}
    where $\rho^{(i)}=\sqrt{\langle\widehat{\mathcal{H}}^{(i)},\widehat{\mathcal{H}}^{(i)}\rangle}$, with the standard inner product expressed as follows
    \begin{equation}
        \langle a,\,b\rangle=4\int_{f_\mathrm{lower}}^{f_\mathrm{upper}}\frac{\hat{a}(f)\hat{b}^\ast(f)+\hat{a}^\ast(f)\hat{b}(f)}{2}\frac{\mathrm{d}f}{S_h(f)}\;.
    \end{equation}
        The noise power spectral density of the ET is denoted by $S_h(f)$, and is illustrated in Figure~\ref{fig:noises}. The upper cutoff frequency is dictated by the last stable orbit of the binary system \cite{6}, while the lower cutoff frequency is set to $1\,\mathrm{Hz}$. Following the adopted SNR threshold for the current GW detectors, we consider a GW detection if the three ET interferometers have a network SNR of $\rho_\mathrm{net}>8$. Assuming that the error on $D_{L,\mathrm{GW}}$ is uncorrelated with any other GW parameter, we can estimate the instrumental error via a Fisher information matrix, leading to the following expression
    \begin{equation}
        \sigma_I\simeq\sqrt{\Bigg\langle\frac{\partial\widehat{\mathcal{H}}}{\partial D_{L,\mathrm{GW}}},\frac{\partial\widehat{\mathcal{H}}}{\partial D_{L,\mathrm{GW}}}\Bigg\rangle^{-1}}\;.
    \end{equation}
    Moreover, since $\widehat{\mathcal{H}}\propto D_{L,\mathrm{GW}}^{-1}$, we arrive at $\sigma_I\simeq2 D_{L,\mathrm{GW}}/\rho$, where the factor of two was introduced in order to take into account the maximal effect of the binary's inclination angle on the SNR. We should also remark that one could adopt the following fitting function for the projected instrumental error contribution of the ET to the relative error on the luminosity distance measurement \cite{6}
    \begin{equation}
        \sigma_{I} =0.1449z-0.0118z^2+0.0012z^3\;.
    \end{equation}
    \item \textbf{Weak lensing error :} It is introduced, since standard sirens get lensed in identical fashion to EM sources, the inhomogeneities along the line--of--sight give rise to a weak lensing effect. In the weak lensing regime, the magnification $\mu_{WL}$  can be expressed to first order in terms of the convergence $\kappa$ as 
    \begin{equation}
        \mu_{WL} \simeq 1+2\kappa\;.
    \end{equation}
    Therefore, we will adopt the following weak lensing uncertainty \cite{wlerror} 
    \begin{equation}
        \sigma_{WL}=\frac{0.1z}{1 + 0.07z}\;.
    \end{equation}
    \item \textbf{Peculiar velocity error :} The peculiar velocity of the source relative to the Hubble flow introduces another additional error. We consider the following functional form for this error \cite{18}
    \begin{equation}
        \sigma_P^2 = \left[1+ \frac{c(1+z)^2}{H(z)D_L}\right]^2\frac{\langle v^2\rangle}{c^2}\;,
    \end{equation}
    where we assume a r.m.s. velocity of $\langle v \rangle=500 \ \rm km/s$ based on numerical simulation results from \cite{21}.
\end{itemize}
We are now in a position to write down the Fisher matrix for the cosmological parameters of our given model, which can be expressed as follows
\begin{equation}
		F_{ij} = \sum_{n=1}^{1000} \frac{1}{(\sigma^2)_n} \left.\frac{\partial D_L(z_n)}{\partial\theta_i}\right|_{\rm fid} \left.\frac{\partial D_L(z_n)}{\partial\theta_j}\right|_{\rm fid}\;,
		\label{fisher}
\end{equation}
where the sum runs over all the $1000$ standard siren events. The partial derivatives of $D_L$ (equation (\ref{h0})) are with respect to the cosmological parameters $\Theta=\{\Omega_m^0,\,H_0^{},\,w_0^{},\,w_a^{}\}$, computed at their fiducial values $\Theta_\mathrm{fid}=\{0.315,\,67.4,\,-1,\,0\}$, adopted from the latest CMB inferred constraints \cite{Aghanim:2018eyx} in the $\Lambda$CDM framework.

\begin{figure}
    \includegraphics[width=\columnwidth]{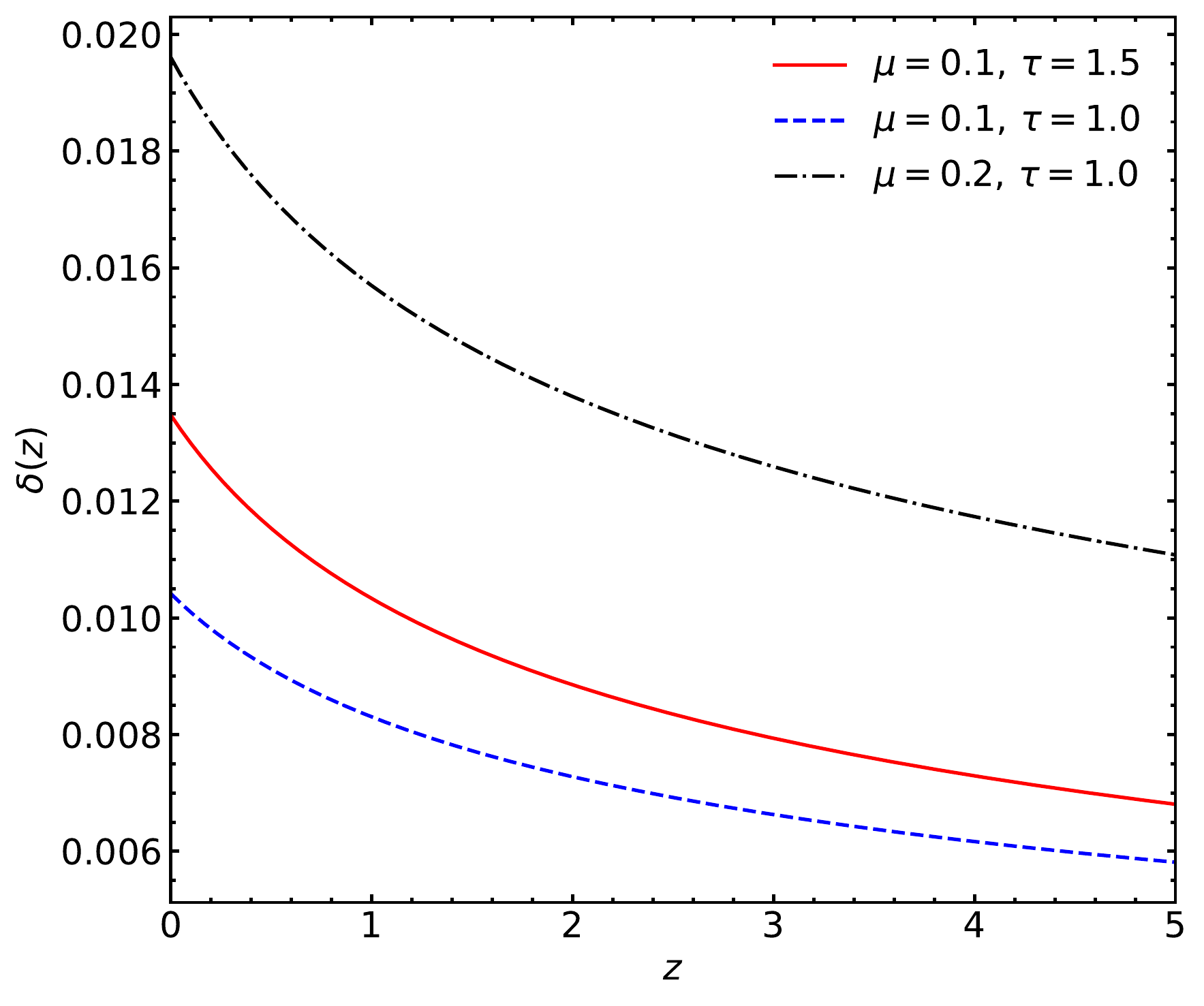}
    \caption{The redshift evolution of the $\delta(z)$ function, for the specified parameters as indicated in Table~\ref{t0}.} 
    \label{delta}
\end{figure} 
\begin{figure*}
    \includegraphics[width=\textwidth]{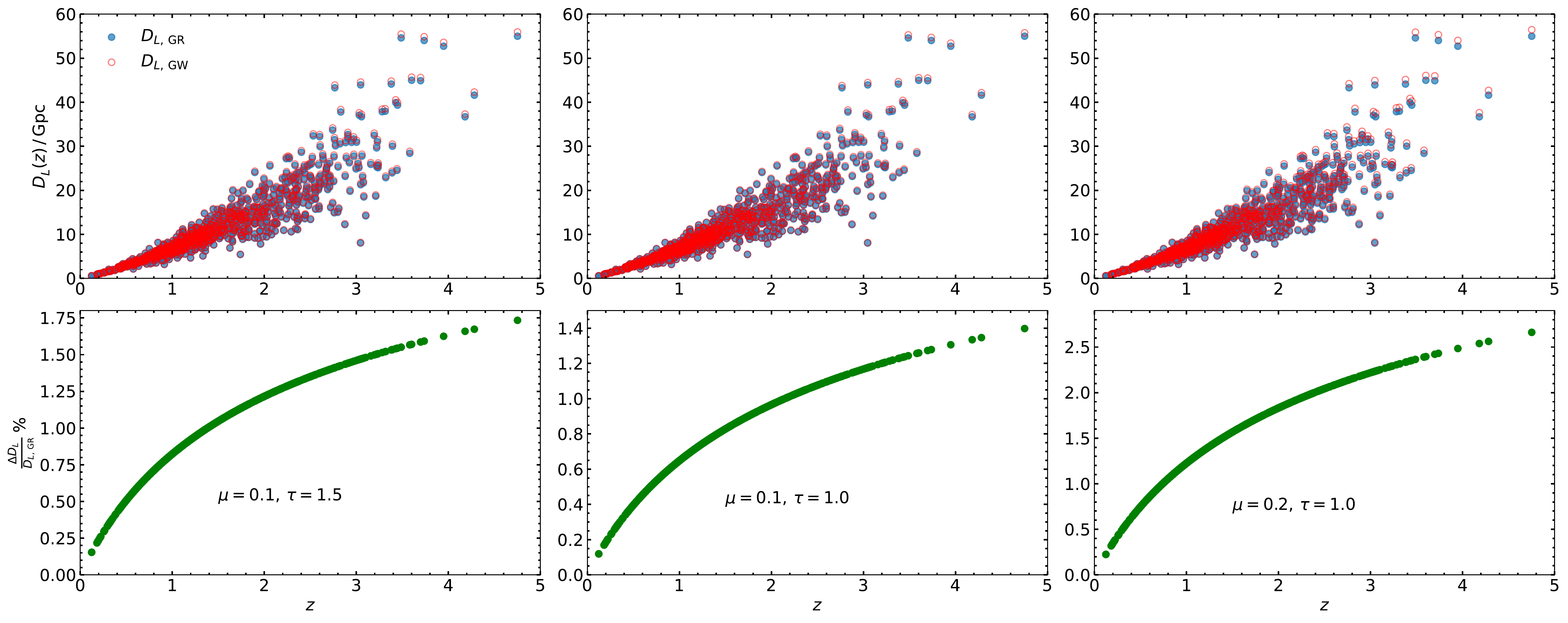}
    \caption{Comparison between $D_{L,\,\mathrm{GW}}$ and $D_{L,\,\mathrm{GR}}$, where the former is computed via equation (\ref{e4}). The lower panel shows the relative error between the two luminosity distances.}
    \label{f5}
\end{figure*}
%

\section{Results}
\label{sec:res}

As mentioned above, for our Horndeski model analysis, we used the No Slip Gravity model. For the Fisher analysis, we have to choose the parametric results of the model parameters. Precisely, we needed the value of the parameters $(\zeta_0,\,n)$ and for that we also needed the value of $\alpha_{M}^0$ (equations (\ref{z0}, \ref{z1})). It is evident that for the computation of the above parameters, via equations (\ref{9}, \ref{10}, \ref{delta2}), we need to know $\tau,\,\mu$ and $a_t$. \cite{5} has presented three different sets of viable parameter values of $\tau$ and $\mu$ for a fixed $a_t=0.5$, which we summarise in Table~\ref{t0}. By adopting these parameter sets, we computed three respective sets of $\alpha_{M}^0$ along with the corresponding values of $\zeta_0$ and $n$ using equation (\ref{z0}) together with equation (\ref{z1}). The derived values are summarised in Table~\ref{T1}. 

It is interesting to note that \cite{16} proposed fit values for the above parameters using another alternate modified gravity model, the \textit{RR} model, which is specified by
\begin{equation}\label{rr}
    [\zeta_0,\,n] =[0.970,\,2.5]\;.
\end{equation}
In appendix~\ref{A1}, we further present a comparison of our results with this model parametrisation.

\subsection{Luminosity Distance}

In this section, we will present a comparison of the computed luminosity distances ($D_{L,\,\mathrm{GW}}$) using the parameter values as specified in Tables \ref{t0}--\ref{T1} with the luminosity distances from Einstein gravity ($D_{L,\,\mathrm{GR}}$). Based on equation (\ref{e4}), we calculated $\delta (z)$ from the parameter sets outlined in the previously mentioned tables. The different redshift evolution of $\delta (z)$ for different models are shown in Figure \ref{delta}. 

Moreover, Figure~\ref{f5} shows a comparison of the distance estimates from three sets of parameter values. One could observe that distance estimates are more sensitive to the amplitude of transition parameter $\mu$, compared to the rapidity $\tau$.

\subsection{Fisher Analysis}
\label{sec:fisher}

\begin{figure*}[htp]
\includegraphics[width = 0.99\columnwidth]{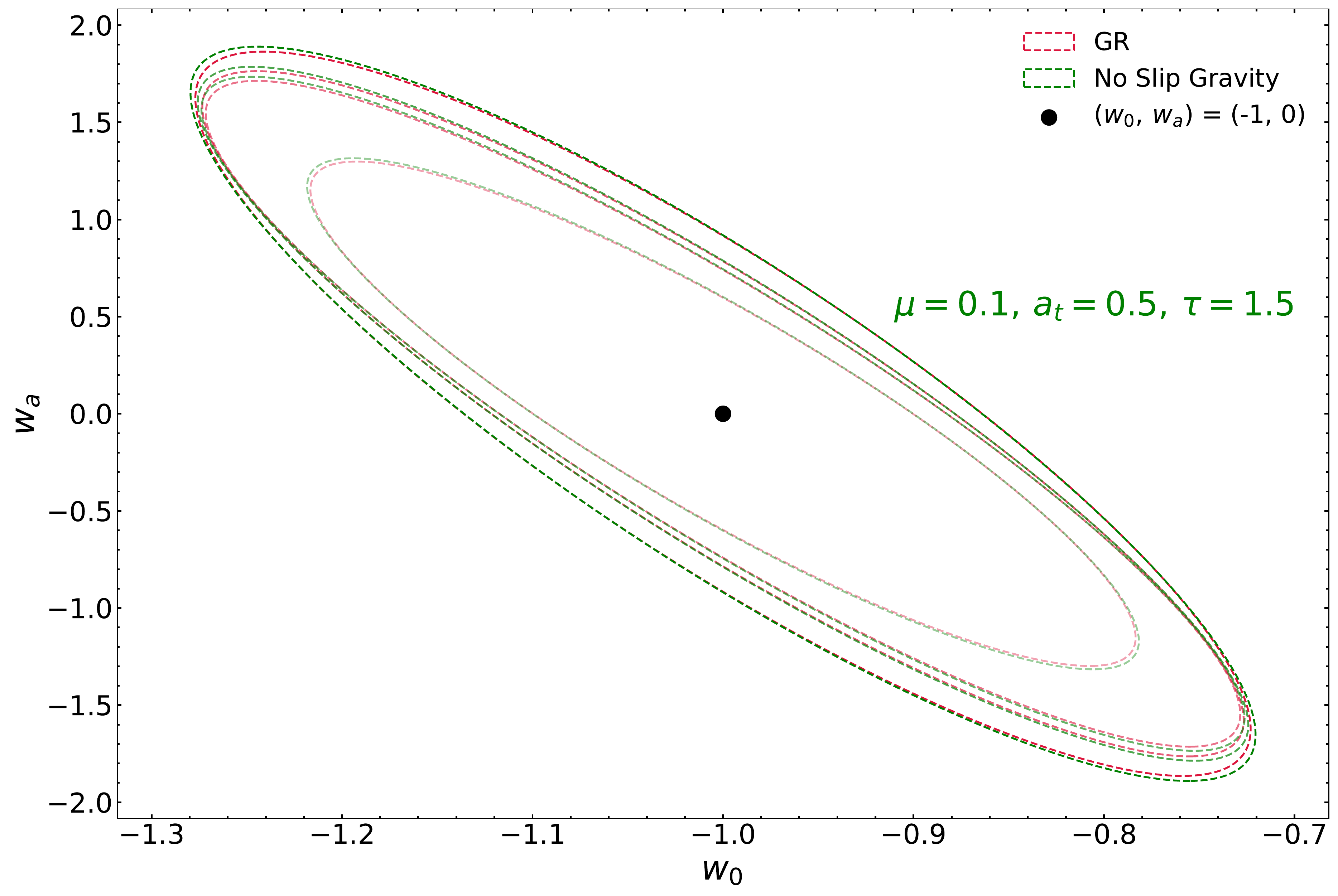}
\includegraphics[width = 0.99\columnwidth]{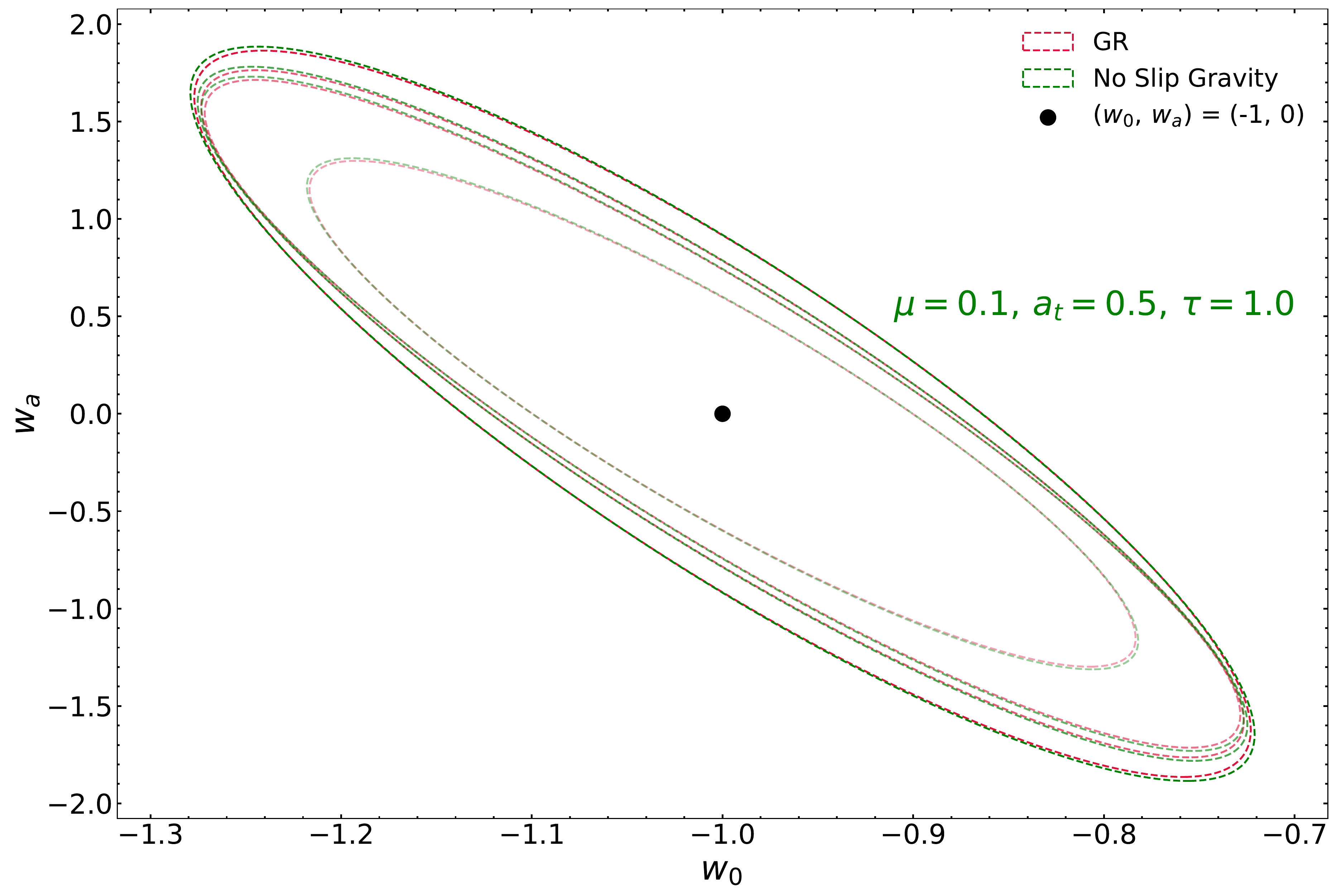}
\includegraphics[width = 0.99\columnwidth]{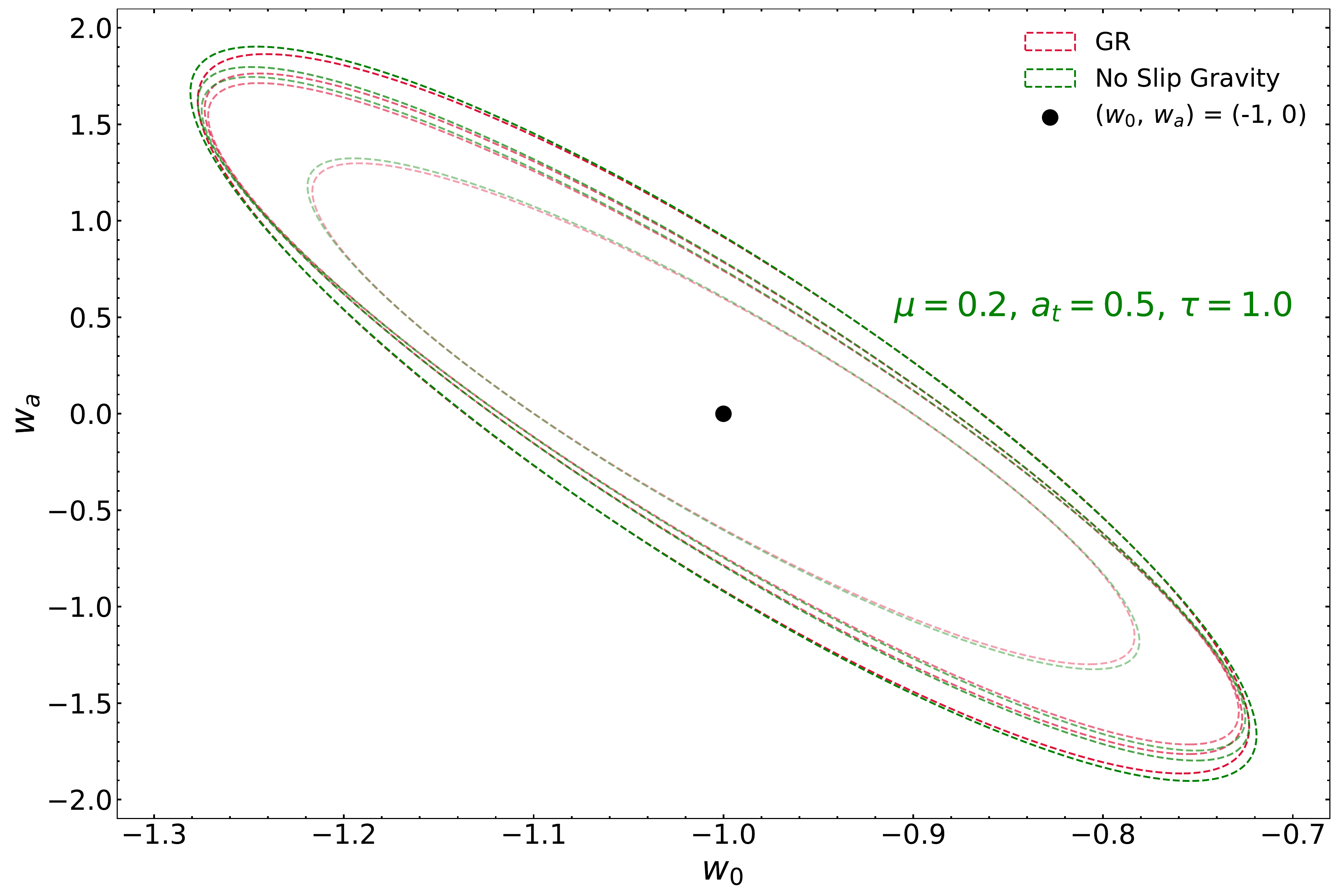}
   \caption{Fisher ellipses computed by using the luminosity distances $D_{L,\,\mathrm{GW}}$ for each respective parameter set in the No Slip Gravity model (green), compared with the ellipse derived from $D_{L,\,\mathrm{GR}}$ (red). Each panel has a set of four concentric ellipse pairs, which correspond to the $z_\mathrm{spectro}$ coverage from $[0.2-4.0]$ as we traverse from outside to inside.}
   \label{fig:default}
\end{figure*}

In \cite{5} the combination of $\left[ \mu,\,\tau,\,a_t\right] = \left[ 0.1,\,0.5,\,1.5\right]$ was found to be in good agreement with current observations. In this section we will present a comparison of the $1-\sigma$ constraint plot (ellipses)  on the dark energy equation of state parameters $w_0-w_a$ under the assumption of all three previously defined parameter sets. In  Figure~\ref{fig:default} the three different parameter choices are considered, with each of them being analysed in four distinct spectroscopic redshift availability criteria. Three of the four spectroscopic redshift coverage ranges are chosen as $z_\mathrm{spectro}=[0.2,\,0.3,\,0.5]$, and as a hypothetical benchmark result we  showed what is the constraint if there was a (hypothetical) full spectroscopic redshift coverage. At the time of writing this paper, our best guess is to assume that up to redshift of $z\simeq0.3$, there will be possible spectroscopic observations and thus the availability of the spectroscopic redshift. The concentric ellipses in each plot, from outside to inside, cover  $z_\mathrm{spectro}=[0.3-4.0]$ cases. Each of these cases are plotted in pairs of ellipses from the No Slip Gravity model and the corresponding Einstein gravity model. We can clearly notice that there is a degeneracy in our results between the three parameter combinations. In all considered cases, the No Slip Gravity model is found to be closely related with Einstein gravity inferred results. As expected, the constraints get tighter as we increase the spectroscopic coverage.    

\begin{figure}
    \includegraphics[width=\columnwidth]{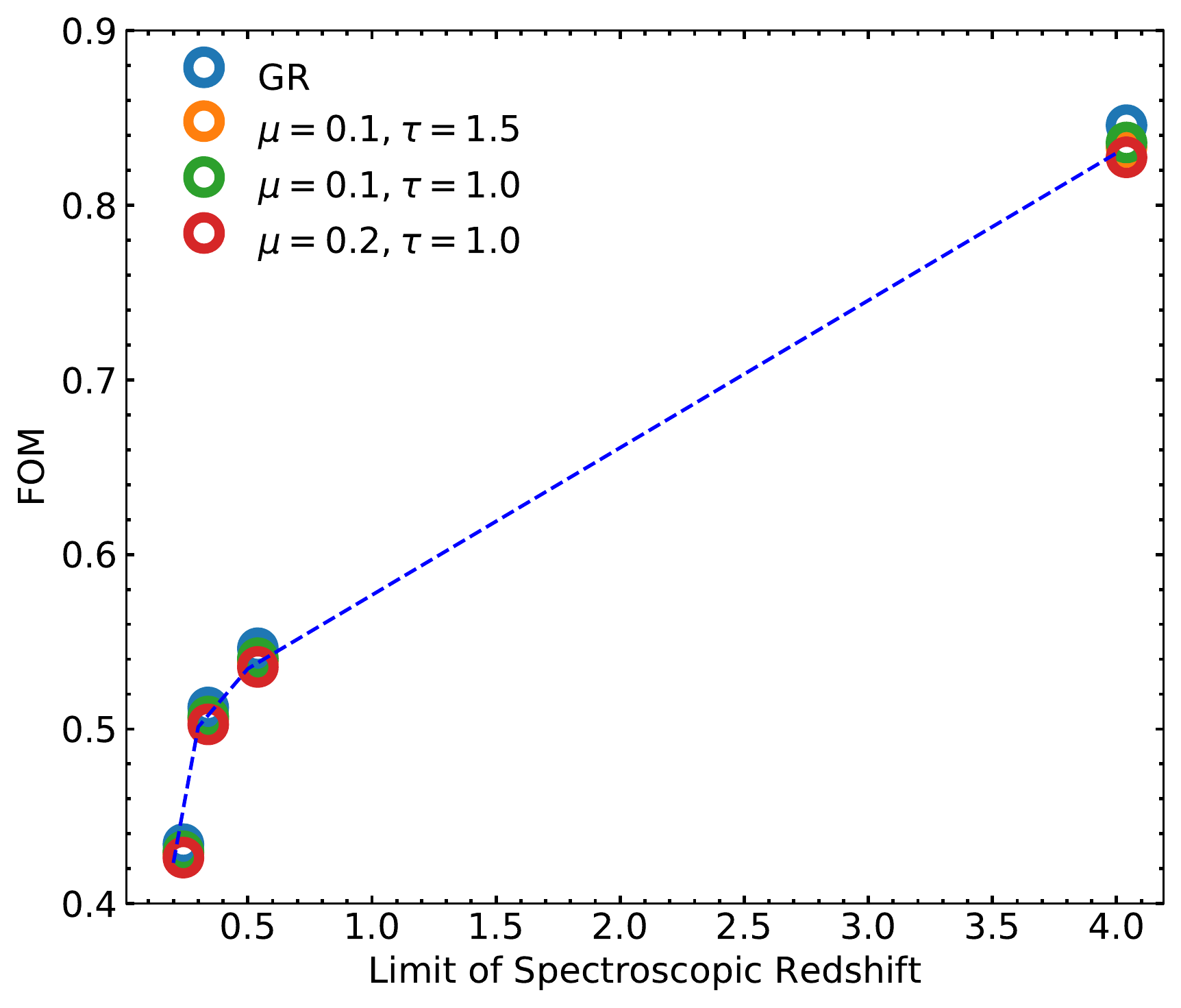}
    \caption{The FOM of the confidence regions shown in Figure~\ref{fig:default}  (Table~\ref{T13}) as a function of the spectroscopic redshift coverage. The blue dashed line corresponds to the mean FOM of the four values at each $ z_\mathrm{spectro}$. }
    \label{fom}
\end{figure}

A measure of these constraints can be analysed by using the Figure of Merit (FOM) values \cite{28}. In Table~\ref{T13} we have summarised the FOM values of the ellipses found in Figure~\ref{fig:default}, and these are further illustrated in Figure~\ref{fom}. There is consistency in the trend of the FOM values throughout the three parameter combinations as a function of the spectroscopic redshift coverage. We also observe that there is a remarkable improvement in the FOM as we increase the spectroscopic redshift from $0.2$ to $0.3$. In contrast, the improvement in the FOM for the remaining length of the abscissa is less steeper. Definitely, at the time of ET's realisation, we expect this redshift range to be spectroscopically covered. The blue dashed line gives a locus of the FOM as a function of the spectroscopic redshift coverage, all the way to full spectroscopic redshift availability. Significant improvements can be made by focusing on the enhancement of spectroscopic redshift availability in the low--redshift ranges. We should further remark that there is no significant model sensitivity in terms of the FOM, although visibly minuscule deviations could be noticed at the highest redshift coverage between  Einstein gravity and the No Slip Gravity models.    

\begin{table}
    \setlength\extrarowheight{2.09pt}
    \begin{center}
        \begin{tabular}{ c c c c }
            \hline
                $z_{\rm spectro}$ &     FOM &  Set \\[.3em]
            \hline
              0.2 &  0.4283 &  GR \\
              0.3 &  0.5066 &   \\
              0.5 &  0.5405 &   \\\hline
              0.2 &  0.4231 &  NSG-I \\
              0.3 &  0.5000 &   \\
              0.5 &  0.5332 &   \\\hline
              0.2 &  0.4242 &  NSG-II \\
              0.3 &  0.5014 &   \\
              0.5 &  0.5348 &   \\\hline
              0.2 &  0.4206 &  NSG-III \\
              0.3 &  0.4969 &   \\
              0.5 &  0.5298 &   \\
            \hline
        \end{tabular}
    \end{center}
    \caption{Summary of the FOM values of the plots shown in Figure~\ref{fig:default}, as a function of the spectroscopic redshift availability ($z_{\rm spectro}$) excluding the case of full spectroscopy, i.e. $[0.2,\,0.3,\,0.5]$. The third column corresponds to either Einstein gravity (GR) or the adopted No Slip Gravity (NSG) parameter values, such that  $\rm{NSG-I}:(\mu=0.1,\,\tau=1.5),\  \rm{NSG-II}:(\mu=0.1,\,\tau=1.0)$ and $\rm{NSG-III}:(\mu=0.2,\,\tau=1.0)$. A visual representation is depicted in Figure~\ref{fom}.}
    \label{T13}
\end{table}

\subsection{MCMC analysis}

We also conducted a Markov Chain Monte Carlo (MCMC) analysis with the forecast data, to determine how well the parameters $\zeta_0$ and $n$ could be measured with this data, and what degeneracies might exist with the cosmological parameters that control cosmic distances.

\begin{figure*}
    \includegraphics[width=0.8\paperwidth]{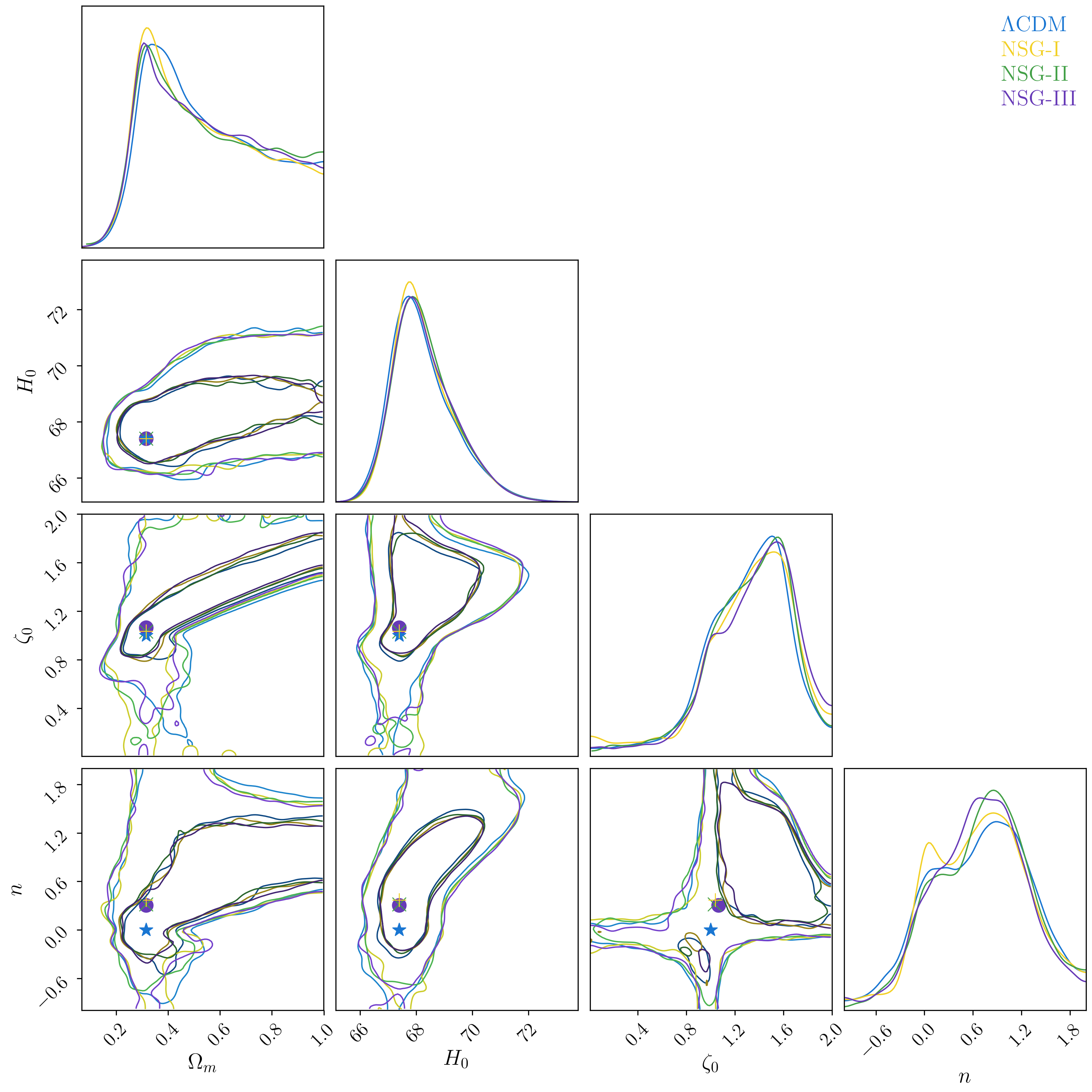}
    \caption{The forecasted 68\% and 95\% Bayesian credible intervals for the parameters $H_0$, $\Omega_m^0$, $\zeta_0$ and $n$, using the prospective data described in section \ref{sec:data}, for each of the four different underlying models ($\Lambda$CDM where gravity is the Einstein model, and three No Slip Gravity models with different choices of $\zeta_0$ and $n$). The bounds are estimated using MCMC, and the different markers represent the values of the parameters for the true model, in the different cases. The bounds on the No Slip gravity parameters are much larger than the differences in the true values of $\zeta_0$ and $n$, and there are significant parameter degeneracies between $\Omega_m^0$ and the No Slip gravity parameters. The constraints are nearly identical between the models using this data set.  }
    \label{mcmc_zeta0_n}
\end{figure*}

We use the affine-invariant ensemble sampler for MCMC \cite{GoodmanWeare}, applying the specific Python implementation \verb!emcee! \cite{2013PASP..125..306F}. We adopted uniform priors, with $0 < \Omega_m^0 < 1$, $0 < \zeta_0 < 2 $, and $ -1 < n < 2$. We assume a flat universe and hold all other parameters fixed to their fiducial values, including the dark energy parameters $w_0=-1$ and $w_a=0$. We ran separate MCMC analyses for each of the four models, to determine if the data could distinguish between the models (assuming each of them to be true).  The Bayesian credible contours are shown in Figure \ref{mcmc_zeta0_n}, and were generated using \verb!ChainConsumer! \cite{2016JOSS....1...45H}.

We find that the data is not constraining enough to be able to to distinguish between the different models. The difference between values of $\zeta_0$ and $n$ given in Table \ref{T1} is much smaller than the parameter bounds, and so the models are indistinguishable. The constraints on these parameters are also identical, confirming the results from section \ref{sec:fisher}.

We see a significant degeneracy between the No Slip gravity model parameters $\zeta_0$ and $n$, and the matter density $\Omega_m^0$. As the matter density increases, the luminosity distance to the different GW sources will decrease, but this can be balanced by increasing the amount of GW `dimming' that is generated by the modified gravity model, making the sources appear to be further away. This degeneracy is also present between the No Slip gravity parameters and $H_0$, though to a lesser extent. Since the Hubble parameter is mainly constrained by the data at low redshift, the amount of distance available to see a significant impact on the value of $H_0$ is reduced.  This is why the values of $n$ and $\zeta_0$ need to be quite large before the Hubble parameter is significantly shifted.

Though the model cannot be distinguished from Einstein gravity using this data by itself, it may be possible to do so in combination with other distance probes. Since luminosity distances (and angular diameter distances) measured by electromagnetic means (for example, SNe Ia or BAO) will be completely insensitive to the No Slip gravity model parameters, they can provide independent constraints on the cosmological parameters $H_0$ and $\Omega_m^0$. By combining these electromagnetic data sets with the GW luminosity distances, the degeneracy between the cosmology parameters $\Omega_m^0$ and $H_0$ with the No Slip gravity parameters $\zeta_0$ and $n$ can be broken, and the size of the confidence contour can be significantly reduced. This is analogous to testing Etherington's distance-duality equation for electromagnetic distances, with the alteration that here the check is the consistency between electromagnetic and GW distances. We leave such a demonstration for future work. 

\section{Conclusion}
\label{sec:con}

We have presented a study of what we can expect in terms of cosmological analysis from prospective GW detections from the perspective of modified gravity models, particularly focusing on a subclass of the Horndeski scalar--tensor theory of gravitation. We have presented a comparison to compare the cases between a modified gravity model and the standard Einstein gravity. For the implementation of the modified gravity model, we used the No Slip Gravity model, as outlined in \cite{5}. This is primarily motivated from the fact that current observational probes, including GW detections, are in agreement with the predictions of the No Slip Gravity model. Third generation interferometric surveys are projected to be operational post $2030$ and we expect that modified gravity models, including the one analysed here, will be robustly tested by a number of proposed surveys such as LSST \cite{2009arXiv0912.0201L} and Euclid \cite{2011arXiv1110.3193L}. 

From our results, we see that the alternative model mimics the standard Einstein theory for the homogeneous expansion. We find, for the models explored here, that the effect on both the distances measured, and the values of the cosmological parameters recovered,  are small. We show that, considering a GW-only data set, there will be significant parameter degeneracies between the cosmological parameters, such as $\Omega_m^0$, with the parameters of the No Slip gravity model. This is because the `dimming' of the GW luminosity distance can also be mimicked by the change in propagation of the GW in the modified gravity theory. Such a degeneracy could be broken through combining the GW dataset with distances estimated through electromagnetic means, or else through separate constraints on the No Slip gravity model parameters.

We await with great expectations from future GW surveys for demystifying the fabric of gravity and the implications it will have on improving our understanding of precision cosmology.     

\section{Acknowledgement}

We are grateful to the authors of \cite{17} for providing their mock data set. AM acknowledges the support of Orau Grant No. 110119FD4534.  JM would like to acknowledge funding support Cosmology@MALTA which is supported by the University of Malta. DFM thanks the Research Council of Norway for their support and the UNINETT Sigma2 -- the National Infrastructure for High Performance Computing and Data Storage in Norway.

\appendix

\section{Comparison with the \textit{RR} Model.}  \label{A1}

\cite{16} have presented an alternate parametrisation for a different modified gravity model, the \textit{RR} model (for a detailed study, refer to \cite{16,22}). 

\begin{figure}
    \includegraphics[width=\columnwidth]{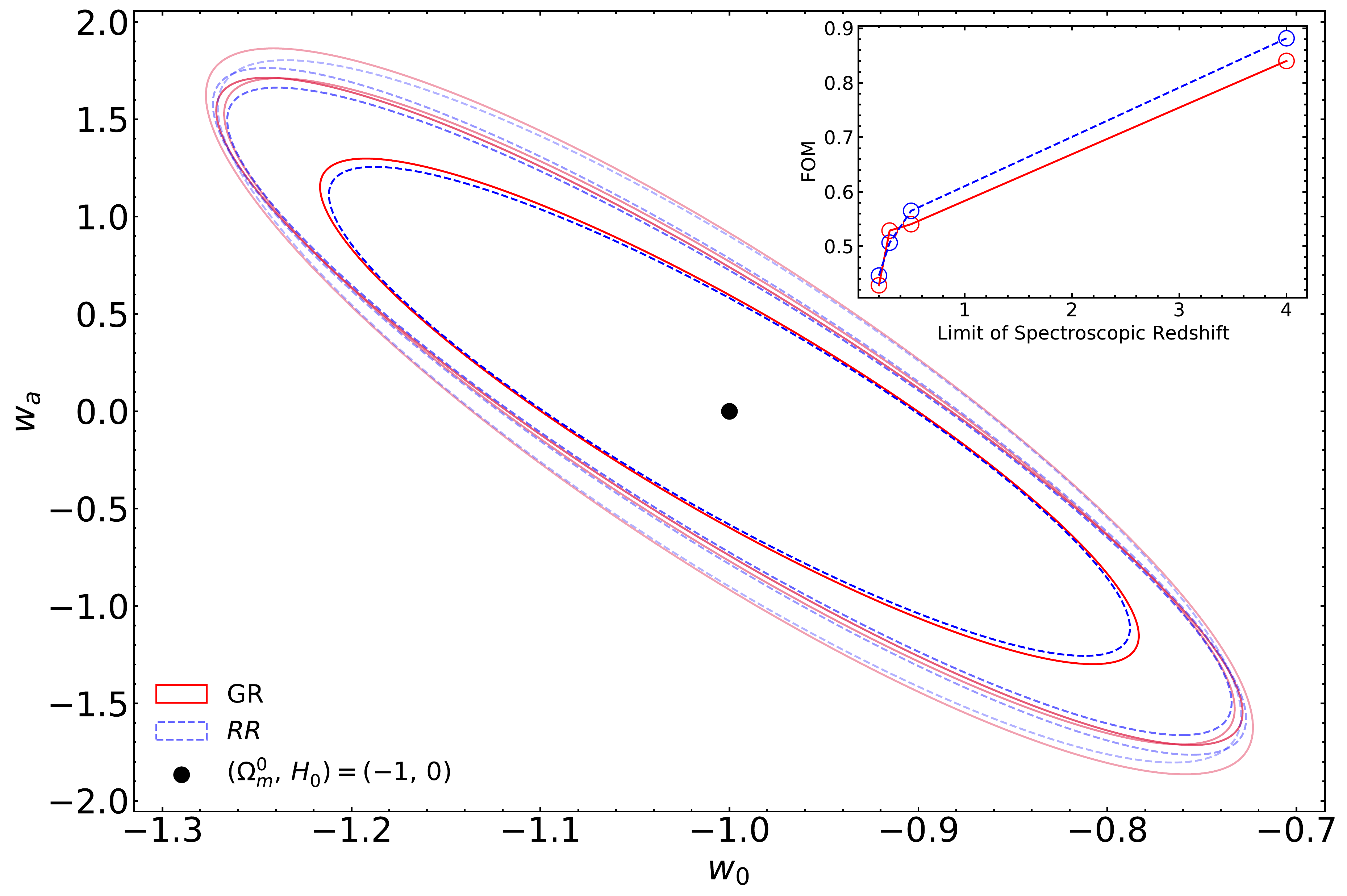}
    \caption{Constraint plots for the \textit{RR} Gravity model (blue) in comparison with the Einstein gravity confidence regions (GR, red) for dark energy parameters $(w_0,\,w_a)$. The concentric ellipse pairs are plotted similar to those in Figure \ref{fig:default}, where  we used the assumption that there will be spectroscopic redshifts available up to $z=[0.2,\,0.3,\,0.5,\,4.0]$ (outside to inside). Inset plot: FOM plot of the corresponding ellipses.}
    \label{a22}
\end{figure}

Although these models, are mostly screened from observational requirements, still for comparisons and reference purposes, we show the similar $1-\sigma$ dark energy parameter constraint plot from this model, with $[\zeta_0,\,n] =[0.970,\,2.5]$ (equation \ref{rr}). It is interesting to see that the \textit{RR} model gives rise a constraint which is marginally tighter than the corresponding Einstein gravity model. 
A measure of their corresponding FOM is also provided in the inset plot of Figure~\ref{a22}.

\section{Hubble Parameter Constraints with the No Slip Gravity Model}

We here present the $H_0^{}-\Omega_m^{0}$ constraint figures. The concentric pairs of ellipses are plotted similar to those in Figure \ref{fig:default} as a function of the spectroscopic redshift availability [$0.2$ (second from the outermost), $0.3$, $0.5$, $4.0$ (innermost)]. For No Slip Gravity we  used the first parameter choice of $\left[ \mu,\,\tau,\,a_t\right] = \left[ 0.1,\,0.5,\,1.5\right]$. The outermost light blue ellipse is a reference showing the constraint if no spectroscopic  redshifts are available.    

\begin{figure}
    \includegraphics[width=\columnwidth]{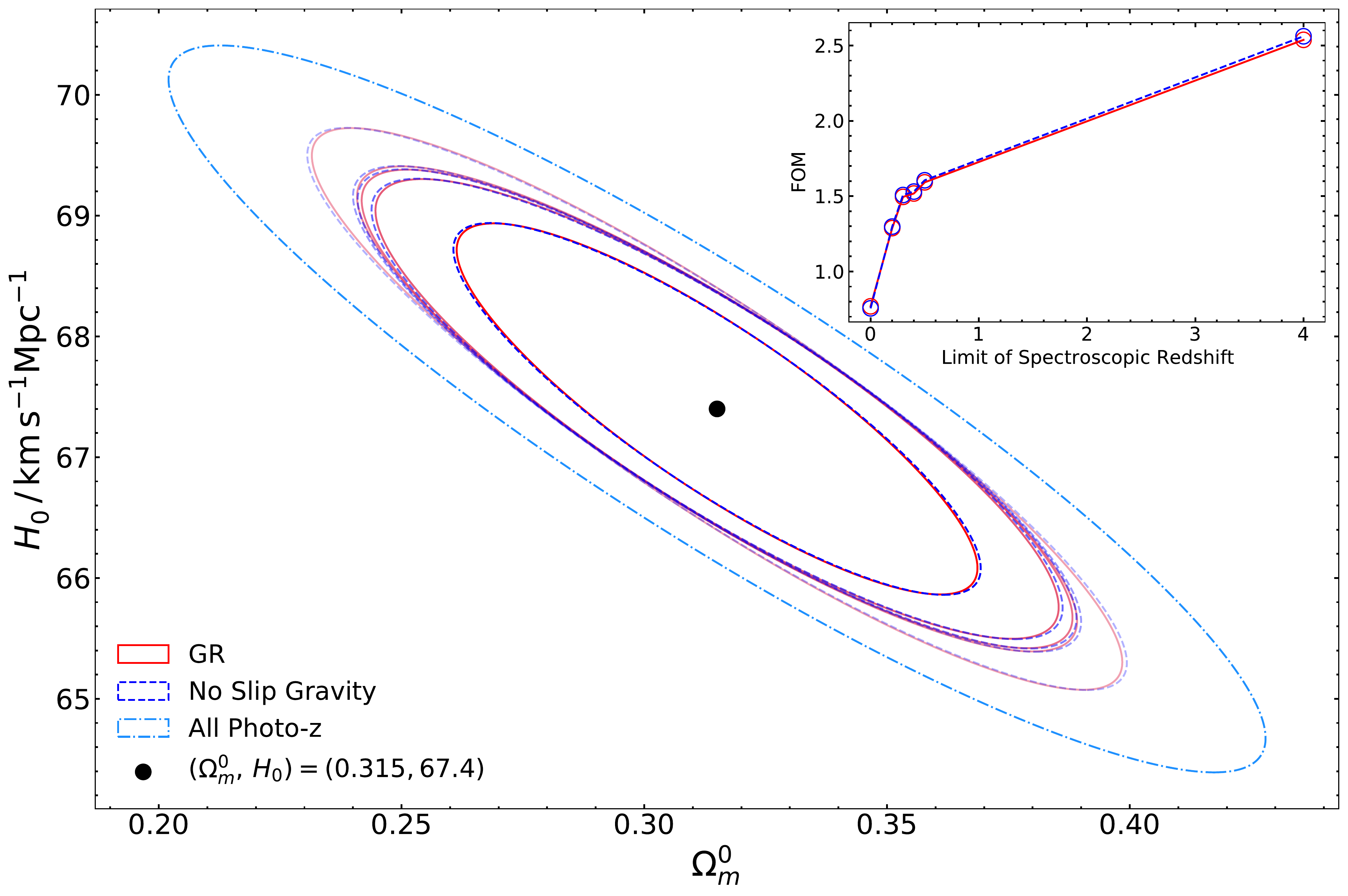}
    \caption{Comparison plot showing the extreme limits of the constraints on the Hubble constant ($H_0^{}$) and the matter density ($\Omega_m^{0}$) plane using the first parameter set (blue dashed) of the No Slip Gravity model from Table~\ref{t0} along with Einstein gravity (GR) (red dashed). As we move from the outer edge to the inner edge, the spectroscopic coverage changes from $[0.2-4.0]$, and the dot--dashed blue confidence region corresponds to the scenario if all redshift was photometric (using GR). Inset plot: FOM of the corresponding ellipses, also listed in Table~\ref{T4}. }
    \label{a3}
\end{figure}

Similar to the observations on the $w_0-w_a$ confidence regions, we see that huge improvements in parameter constraints can be achieved  by using spectroscopic redshifts in the low--redshift range. Indeed, the inset plot of the FOM from Figure \ref{a3} (or Table \ref{T13}) shows this trend. We also see that the modified gravity model closely mimics the standard prediction and that they are nearly identical. Again, we would like to remark that the case of full spectroscopic redshift availability is a hypothetical reference point. This can be thought of as the maximal 
constraint that these parameter pairs can achieve with the given specifications. 

\begin{table}
    \setlength\extrarowheight{2.09pt}
    \begin{center}
        \begin{tabular}{ c c c c }
            \hline
            $\rm z_{spectro}$ & $\rm GR$ & $\rm NSG-I$ \\[.3em]
            \hline
            0.0 & 0.7555 & 0.7689  \\ 
            0.2 & 1.2981 & 1.2880 \\ 
            0.3 & 1.5077 & 1.4953 \\
            0.5 & 1.6055 & 1.5918 \\
            4.0 & 2.5614 & 2.5377 \\
            \hline
        \end{tabular}
    \end{center}
    \caption{Table showing the FOM comparisons between   GR and the No Slip Gravity model from Figure \ref{a3}. The second column is similar in abbreviation to Table~\ref{T13}. }
    \label{T4}
\end{table}

\bibliographystyle{unsrt}

\bibliography{reference2} 

\end{document}